\documentclass[sn-mathphys,Numbered]{sn-jnl}

\usepackage{graphicx}%
\usepackage{multirow}%
\usepackage{amsmath,amssymb,amsfonts}%
\usepackage{amsthm}%
\usepackage{mathrsfs}%
\usepackage[title]{appendix}%
\usepackage{xcolor}%
\usepackage{textcomp}%
\usepackage{manyfoot}%
\usepackage{booktabs}%
\usepackage{algorithm}%
\usepackage{algorithmicx}%
\usepackage{algpseudocode}%
\usepackage{listings}%

\theoremstyle{thmstyleone}%
\theoremstyle{thmstyletwo}%

\theoremstyle{thmstylethree}%

\begin{document}

\title[Article Title]{ Cosmological significance of the early bright galaxies observed with JWST}


\author*[1,2]{\fnm{Moncy V.} \sur{ John}}\email{moncyjohn@yahoo.co.uk}

\author[3]{\fnm{K.} \sur{Babu Joseph}}\email{bb.jsph@gmail.com}

\affil*[1]{\orgdiv{School of Pure and Applied Physics}, \orgname{Mahatma Gandhi University}, \orgaddress{\street{} \city{Kottayam}, \postcode{686560}, \state{Kerala}, \country{India}}}

\affil[2]{\orgdiv{Department of Physics}, \orgname{St. Thomas College}, \orgaddress{\street{} \city{Kozhencherri}, \postcode{689641}, \state{Kerala}, \country{India}}}

\affil[3]{\orgdiv{Department of Physics}, \orgname{Cochin University of Science and Technology}, \orgaddress{\street{} \city{Kochi}, \postcode{682022}, \state{Kerala}, \country{India}}}


\abstract{The  recent discovery of objects with redshift $z>10$ with the help of James Webb Space Telescope (JWST)  poses  serious challenges to the  $\Lambda$CDM cosmological model, which has been in vogue for some time now. The new data indicate that galaxy formation must have taken place much earlier than expected in this model. Another viable class of cosmological models is that of the so-called coasting models, in which the scale factor of the universe varies proportionately with time. In these models, the universe at redshift $z=12$ has ample time ($\sim 1070$ Myrs) for galaxy formation. The earliest such model is the one proposed by E.A. Milne, based on his `kinematic relativity', but it is considered unrealistic for not treating gravity as relevant at cosmological scales. A closed version of an eternal coasting FLRW model was proposed by the present authors even before SNe Ia data began to pour in. Subsequently we developed a more general model of the same class, which is valid for all the three possible geometries, with open, closed or flat spatial sections. In the nonrelativistic era, this model makes the falsifiable prediction that the ratio of matter density to dark energy density is 2. This avoids the cosmic coincidence problem. Moreover, this eternal coasting model allows room for creation of matter from dark energy, that may speed up galaxy and structure formation at the early epochs, as implied by the JWST data. The paper also attempts to review some similar coasting models, but emphasizes the eternal coasting cosmology as the most probable candidate model capable of explaining the presence of  high redshift galaxies discovered by JWST.}

\keywords{Cosmology: Observations, Cosmology: Theory, Early bright galaxies, Coasting cosmological model}



\maketitle
\section{Unexpected early bright galaxies}

The long awaited glimpse of the cosmic dawn at $z>10$ with the James Webb Space Telescope (JWST) has been revealing. The near infra-red camera (NIRCam) imaging data  from the Glass-JWST Early Release Science Program (JWST-ERS-1324) helped to identify several massive and bright galaxy candidates in this till now uncharted cosmic territory. Two particularly luminous sources GL-z10 and GL-z12 were identified by Naidu et.al. \cite{naidu} at redshifts  $z=10.4^{+0.4}_{-0.5}$ and $z=12.4^{+0.1}_{-0.3}$. Castellano et. al. \cite{castellano} measured these objects to be at $z=10.6$ and $z=12.2$, respectively. These  objects appear to be  over a million solar masses  and might have built-up their masses in only $t< 300 - 400$ Myr after the Big Bang. As per the accounts of the  cold dark matter (CDM) cosmology of the early 1990's, galaxies were formed only at around redshift $z\sim 5$ \cite{peebles}. The currently popular $\Lambda$CDM cosmology pushes this limit to  $z\sim 6-9$, but it is noted that even in this case, the above two high redshift objects with $z>10$ are quite unexpected in the small survey volume of the observation \cite{castellano}. 

Another recent paper \cite{labbe} reports  six other candidate massive galaxies at $7.4 \leq z \leq 9.1$, which are estimated to be within 500-700 Myrs after the Big Bang in the $\Lambda$CDM model. The mass in these galaxy candidates would be be a factor of $\approx 20-1000$ higher than the expected values, and the mass density in the most massive galaxies would exceed the total previously estimated mass density by a factor of 2-5. It is observed that these stellar mass densities are difficult to realize  at these redshifts in the standard $\Lambda$CDM cosmology.

Recent discovery  \cite{peng} of the two companion sources to a strongly lensed galaxy SPT0418-47 at $z=4.225$ is claimed to be a good example of the very early mass build-up and structure formation. This object possesses extra-ordinarily high (solar-like) metallicity at the  age of 1.4 Gyr in the $\Lambda$CDM model, which appears to be in tension with the  theories of galaxy formation.

\section{Attempts to resolve the tension}

An immediate response to the challenges posed by these newly observed galaxy candidates is to place novel constraints on galaxy evolution in the cosmic dawn epoch. It is speculated that the star formation efficiency in the early universe is much higher than the expected values, so that such galaxies began forming potentially earlier than current expectations.  It is widely believed that the new data can be reconciled only through a significant transition in the mode of galaxy formation in the early universe.

While attempting to  modify the theory of galaxy formation  so as to alleviate such tension in the $\Lambda$CDM model, we find it equally important to look for possible alternatives to the  cosmic evolution itself at least in the early epochs. Examining such alternatives is legitimate, especially in the context of similar  modifications  to the  cosmic evolution made  in the past. A clear instance is  the discovery of the accelerated expansion of the present universe. Until the SNe Ia data became available in 1998 \cite{riess,perlmutter}, the  expansion of the universe  in the present epoch was considered to be decelerating. When it was found that the new data is incompatible with a decelerating present universe, the cosmology community did not hesitate to accept a model of the universe which is accelerating  in  the present epoch. In this context, investigating alternative  evolution scenarios in the cosmic dawn that can explain the above observations by JWST is well-motivated.

 Arguing  that  the abundance of early bright galaxies is  in tension with the presently accepted evolution of the universe and that it may provide tight constraints on its expansion history in the early epochs, Menci et. al.  \cite{menci}  puts to test a wide class of dynamical dark energy models, which adopt the dark energy equation of state parameter in the form $w=w_0+w_a(1-a)$. They find, under the light of new observations, that in such a model  a major portion of the parameter space $(w_0,w_a)$ allowed by existing cosmological probes is  ruled out with a confidence level $>2\sigma$.

\section{Eternal coasting universe}

Here we consider an alternative  cosmological model  that can offer potential solution to this impasse by way of having an evolution different from that of the $\Lambda$CDM model. This model is referred to as `eternal coasting', for in it the scale factor of the universe always varies  linearly with time ($a \propto t$). Various aspects of such  models  are already  discussed in the literature.   An extensive review of several linear and quasi-linear coasting models  can be found in Ref. \cite{casado}.

A closed ($k=+1$) version of an eternal coasting model was presented in \cite{mvjkbj1996,mvjkbj1997}, even before the release of SNe Ia data  \cite{riess,perlmutter}. This model was arrived at while investigating a possible `signature change' in the early universe. It is referred to as a `bouncing and coasting' universe\cite{mvj2021}, since it has a prior  contracting phase, has a smooth bounce at a Planck-sized minimum radius $l_p$ and expands linearly thereafter. For $a\gg l_p$, the total matter/energy density $\rho$ and pressure $p$ of the physical universe obeys $\rho+3p=0$, which is called the `zero active gravitational mass' condition. The presence of dark energy makes it a Friedmann-Lamitre-Robertson-Walker (FLRW) model. In \cite{mvjvn2002,mvj2005}, it was shown that this model performs almost equally well when compared to the $\Lambda$CDM model in explaining the SNe Ia data. We find that  the objects discovered in  the Early Release Science Program can have ample age  even at redshifts $z>10$ and hence the model does not encounter any   problem with their discovery.

In this paper we highlight a similar FLRW  model  \cite{mvjkbj2000}, which is more general than  the above one in that it considers all the three possible geometries; i.e., geometries with closed, open and flat ($k=\pm 1,0$) spatial sections. This `eternal coasting' model was proposed on the basis  of some dimensional considerations, which led to the conclusion that  densities of all matter/energy components in the classical universe must vary as the inverse square of the scale factor. In consequence, as in the closed model \cite{mvjkbj1996,mvjkbj1997}, the total energy density   obeys $\rho +3p=0$. Here, the active gravitational mass of the universe must be zero throughout its evolution, starting from a singularity. The cosmic fluid was assumed to contain visible  and dark   matter with density $\rho_m$ and an equation of state $p_m=w\rho_m$ ($w=1/3$ for radiation and $w=0$ for nonrelativistic matter), and also a time-varying dark energy $\rho_{\Lambda}(t)$ with equation of state $p_{\Lambda}=-\rho_{\Lambda}$.  The time evolution of the scale factor is coasting,  given by

\begin{equation}
a=mt,
\end{equation}
where $m=\sqrt {k/(\tilde{\Omega }-1 )}$ is a
proportionality  constant; $\tilde{\Omega }$ is the total density
parameter. With $\Omega_m$ as the density parameter for matter and $\Omega_{\Lambda}$  that for dark energy, we have $\tilde{\Omega}=\Omega_m +\Omega_{\Lambda}$. In the nonrelativistic era we have  $\Omega_m/\Omega_{\Lambda} =2$, where one can write $m=\sqrt {2k/(3\Omega _{m} -2)}$.

At the observational front,   the luminosity distance and the angular diameter distance are the two most important measured quantities  in cosmology. For all  coasting models with  $a=mt$, the luminosity distance is given by the expression \cite{mvjvn2002,mvj2005},

\begin{equation}
D_L^{EC}=\frac{c}{H_0}m(1+z)\sin n \left[ \frac{1}{m}\ln (1+z)\right], \label{eq:luminosityd_coast}
\end{equation}
and the angular diameter distance by

\begin{equation}
d_A^{EC}=\frac{c}{H_0}\frac{m}{(1+z)} \sin n \left[\frac{1}{m} \ln (1+z) \right]. \label{eq:angdimd_coast}
\end{equation}
Here all the three geometries are considered; with  $\sin n (x) = \sinh (x)$ for open,  $\sin n (x) = \sin (x)$ for closed  and  $\sin n (x) = x$ for flat geometries.

In addition to the absence of  horizon, flatness and  other related cosmological problems, the eternal coasting  model has the advantage that it does not have the coincidence problem in cosmology. This latter problem  persists  in the $\Lambda$CDM model, since it is not solvable under inflation.  In the concordance  model, the present values of density parameters are $\Omega_m =0.3$ and $\Omega_{\Lambda}=0.7$. But since  the density of  matter   varies as  $\rho_m\propto a^{-4}$ (relativistic) or $a^{-3}$ (dust)  and the dark energy component  $\rho_{\Lambda}$ (with equation of state $p_{\Lambda}=-\rho_{\Lambda}$) remains a constant, their respective density parameters  in the $\Lambda$CDM model approach $\Omega_m \rightarrow 1$ and $\Omega_{\Lambda} \rightarrow 0$ as $a\rightarrow 0$. For instance, at the epoch of CMB emission, the respective values are $\Omega_m\approx 1-10^{-9}$ and $\Omega_{\Lambda} \approx 10^{-9}$, which of course can be achieved only by some fine tuning. 

 On the contrary, a feature unique to the eternal coasting model is that the density parameters such as $\Omega_m$ and $\Omega_{\Lambda}$ are  constants in time.  A clearly falsifiable prediction of the model is that the ratio between densities of matter and dark energy is 2. It was reported in \cite{mvjkbj1996} that this model predicts $\Omega_m=4/3$ and $\Omega_{\Lambda}=2/3$, but it must be recalled that these are only for a closed universe with $\tilde{\Omega}=2$. Even in this case, if we consider the cosmic fluid as containing some additional cosmic string-like K-matter with equation of state $p_K=-(1/3)\rho_K$ \cite{kolb}, values of these density parameters can be found to be much smaller  than the above values \cite{mvjkbj1997}. In \cite{mvjkbj2000} too, these values are moderate for lower values of $\tilde{\Omega}$. However, in all these cases, the condition $\Omega_m/\Omega_{\Lambda} =2$ remains valid. For instance, when we consider the $k=0$ case of the eternal coasting model, the values $\Omega_m$ and $\Omega_{\Lambda}$ are 2/3 and 1/3, respectively. (One need not compare it with the respective values 0.3 and 0.7  evaluated in the $\Lambda$CDM model, for these values  are   obtained from model-dependent evaluations \cite{mvj2004} that assume $\Lambda$CDM model to be the true one at the outset.) 
 
 For  all cases (closed, open or flat) of the eternal coasting model, the age of the universe at $z=12$ can be seen to be $\sim 1.07$ Gyr, given the present value of Hubble parameter as  70 km s$^{-1}$ Mpc$^{-1}$. Even for larger values of $z$, the age of the universe is substantial; for $z=20$, it is $\sim 700$ Myr. The redshift corresponding to an age 500 Myr can be seen to be $z=27$. These values show that it would not be surprising  if galaxies with redshifts up to such large values of $z$ are detected in future observations. The new  early bright galaxy candidates observed with JWST do not create any problem in this model.
 
 \section{Other linear expansion models}

A coasting expansion valid only for the early epoch in the universe was proposed by Ozer and Taha \cite{ozertaha}. This model coincides with eternal coasting models in \cite{mvjkbj1996,mvjkbj1997}  during the relativistic epoch. After this era,  the Ozer-Taha model deviates from coasting evolution and merges  with the decelerating Friedman-Robertson-Walker (FRW) model.  Another model by Kolb \cite{kolb} has coasting evolution only in the late universe, when it is dominated by the exotic K-matter mentioned above, with equation of state $p_K=-(1/3)\rho_K$. The agreement is only in the late epochs with low $z$ since  in the early epochs of this model, matter/radiation (which varies as $a^{-4}$) must dominate over K-matter (which varies as $a^{-2}$). 

Based on some theory of gravity other than general relativity,  Dev et al \cite{dev}  arrives at a coasting model that is similar to the eternal coasting model in its expansion history. They have studied several features of this model, such as primordial nucleosynthesis, structure formation etc. and established its viability on these fronts. Recently, F. Melia and collaborators  have worked on a general relativistic `always' coasting model under the title `$R_h=ct$' model and published  its notable successes  in a series of papers \cite{melia2007,melia2012,melia2019,melia2019a}.   Specifically, the $R_h=ct$ model  coincides with the flat ($k=1)$ case of the latter \cite{mvj2019}. However, this  coasting  model differs from the one in \cite{mvjkbj2000}  in one important aspect;  though it maintains the condition $\rho+3p =0$ for the the total energy density and pressure, it  does not prescribe any clear evolution of the individual energy components. A fallout  of this is that it cannot offer a natural solution to the coincidence problem. 

 The earliest model of the universe, which is expanding linearly with time, is the one  conceived by Milne \cite{milne},  on the basis of his kinematic relativity. The original Milne model has  Minkowski spacetime as its background. It does not  take into account any role for gravity in the cosmic dynamics and hence is not generally considered  a realistic model.
 Recently an attempt is made to picture  our universe as undergoing a coasting expansion,  by reviving the  Milne model. It is easy to see that its expansion history, characterized by  the luminosity distance and angular diameter distance, is the same as that in the $\tilde{\Omega}=0$ case  of a Friedmann model. Note that the expressions used in \cite{vishwa} to evaluate these quantities are obtained  by putting $k=-1$ and $\tilde{\Omega}=0$  in  the above equations (\ref{eq:luminosityd_coast}) and (\ref{eq:angdimd_coast}), respectively, used in \cite{mvjvn2002}.  The model in \cite{vishwa} too considers gravity as insignificant in cosmic phenomena, even at the early and dense stages of its expansion. This continues to make it an unreal model.
 
 In the light of the new observations by JWST, Melia  \cite{melia2023} notes that the time-line predicted by the $R_h=ct$ model would fit the birth and growth of the high redshift galaxy candidates observed by JWST, in a much better way than that in the $\Lambda$CDM model. In this context, we note that all the three cases in the eternal coasting cosmology, with $k=\pm 1,0$, have the same age at a given redshift $z$ and hence must be investigated together as possible solutions to the new `age problem' at high redshifts.  
 
 \section{The course ahead for the eternal coasting model}
 
The first point to note is that in the eternal coasting model, which is also a dynamical dark energy model,  one needs to consider all the three spatial geometries. In the $\Lambda$CDM model, there  is the  problem of fine tuning of density parameters, which is resolved only with inflation. In fact, inflation is assumed to make the observed universe as exactly  flat and thus circumvents the flatness problem. However, the coincidence problem is  unsolved in it. On the other hand, in the eternal coasting model, there is no  problem of fine tuning of density parameters and hence no flatness problem. Similarly, it predicts a constant ratio between density parameters and hence there is no  problem of their coincidence in the present epoch. Hence in the eternal coasting model,  one can conceive a cosmological evolution without inflation. We may also note that  in the linear coasting models,  there is no justification for fixing $k=0$ at the outset, as done in the $R_h=ct$ model.  

We have seen that a characteristic feature of the eternal coasting model   is that in it the density parameters $\Omega_m$, $\Omega_{\Lambda}$, etc. are constants. This is possible due to the creation of matter from dark energy. In the nonrelativistic era, the creation of matter is at the rate \cite{mvjkbj1997} 

\begin{equation}
\frac{1}{a^3}\frac{d}{dt}(a^3\rho_m)=\rho_m H, \label{eq:creation_rate}
\end{equation}
where $\rho_m$ is the density of matter and $H$ is the Hubble parameter at the given epoch. In terms of the present matter density and Hubble parameter, one can write this quantity as $\rho_{m0}H_0(1+z)^4$. The creation rate for the present epoch can be found to be extremely small, so that any current observation will be unable to detect it. But it may be noted that at earlier epochs, this creation may have a significant influence on several cosmic phenomena. 

In the $\Lambda$CDM  cosmology, it is assumed that the dark energy distribution in the universe is homogeneous.  In  addition, dark energy will have only an extremely small presence in the early epochs of   this cosmological model. It can be seen that the  ratio $\Omega_{\Lambda} / \Omega_m $, whose  value in the $\Lambda$CDM is $\sim 0.7/0.3$ at present,   would be $ \sim 3\times 10^{-3}$ at the redshift $z=12$.  This is why the dark energy is assumed to have little role in either  galaxy formation or in  structure formation  in the $\Lambda$CDM model. On the contrary,  in the relativistic epoch of the eternal coasting model, dark energy and matter have equal densities and they are in the ratio 1:2 in the nonrelativistic epoch. While attempting to  modify the theories of galaxy and structure formation in the eternal coasting model, the substantial presence of dark energy in the early  epochs needs to be taken into account.

Also, in this model,  there shall be inhomogeneities in dark energy  due to the  creation of matter from dark energy. As  seen from equation (\ref{eq:creation_rate}), the rate of such creation is very large at earlier epochs and will have significant influence in early galaxy formation. At the sites of such creation events where the formation of dark matter halos or star formation take place, there may occur a drop of dark energy density. This, in turn, may help the growth of perturbation of matter density. Perhaps this might have speeded up the  formation of early galaxies now observed with JWST at high redshifts.

Another issue is related to the details of the mechanism of creation of matter. One may presume that creation of matter takes place at the sites of galaxy formation, but if this process has some inherent randomness, the expansion of the universe itself may become stochastic. This is clear from the fact that small fluctuations in the creation rate will lead to fluctuations in the densities of matter and dark energy, which in turn will make the total equation of state parameter $w=p/\rho$ a stochastic force term. Consequently, the Hubble parameter itself will be a stochastic parameter. In \cite{csmvjkbj}, this was proposed as the possible mechanism behind the anomalous scatter of the Hubble diagram at high redshifts. The scatter in the Hubble diagram at low redshifts is known to be due to peculiar velocities,  but that at high redshifts is generally considered as arising due to unknown errors in the measurement. It is worth   noting  that the anomalous scatter at large $z$ did not disappear in spite of the refinement in observational techniques achieved in the past. The above work \cite{csmvjkbj}  explains such scatter by considering  stochastic evolution of the density parameter in the early universe  under the assumption of fluctuating $w$ factor and  by developing a Fokker-Planck formalism for the same. 

In \cite{csmvjkbj}, the stochastic approach was made to a general, flat FLRW cosmology. Later, this formalism was revisited  \cite{mvjcskbj} and applied  to the specific case of the eternal coasting model and  an explanation of the anomalous scatter in the Hubble diagram  at high redshifts was provided. The solution of the Fokker-Planck equation in this case helped to show that the scatter due to fluctuating $w$-factor will increase with redshift. In the same work,  the Hubble parameter $H$ at various epochs, marked by  values of redshift $z$, were evaluated and  the half-width of the theoretical probability distribution for $H$ were compared with that obtained from the apparent magnitude-redshift data of type Ia supernovae. The value of the diffusion constant thus found  was seen to be nearly a constant, as expected. Thus the    assumption of stochasticity due to fluctuating creation rate of matter was found capable of explaining the anomalous scatter in the Hubble diagram at high redshifts. 
 
 It is interesting to note that such fluctuations in creation rates can also have a role in structure formation. The currently favored inflationary expansion at $10^{-35}$ s is generally considered as  indispensable in cosmology, for it will lead to quantum fluctuations in the scalar fields, which in turn will act as the seeds for structure formation. However,  alternative scenarios are possible in this case too. In \cite{berera}, it was suggested that stochastic fluctuations arising from various  dissipation in the classical epoch (such as the creation events discussed above) may lead to generation of seeds for density perturbations in the early universe. If viable, this may completely eliminate the need for inflation in cosmology.

\section{Summary}

The discovery of early bright galaxy candidates at $z>10$ hints at the need for revision of the cosmic expansion history in the early epochs. These objects are difficult to realize in the $\sim 300-400$ Myrs after the Big Bang in the $\Lambda$CDM cosmology.  The  way out of this impasse is  some substantial modification in the galaxy formation scenario of the early universe or  a modified evolution of the universe in the early epochs. In the latter case, the eternal coasting cosmological model  offers a natural solution to the above problem, for it has an age of $\sim 1070$ Myr even at a redshift of $z=12$. The high-redshift galaxy candidates are easy to form within this time interval, under the current theories of galaxy formation.

In the past two decades, several authors have arrived at the idea of a universe model having coasting evolution throughout its history. Some of them are based on alternative theories of gravity and some on no gravity. What comes closest to the present general relativistic FLRW eternal coasting model is the $R_h=ct$ model. The latter model is a special (flat) case of the former. However, there is no clear prescription of the variation of energy densities with time in the $R_h=ct$ model. Hence it is not devoid of the cosmic coincidence problem.

Here we have noted that the eternal coasting model avoids the cosmic coincidence problem by way of having a constant ratio between matter and dark energy densities. Moreover, it calls for modifications in the standard galaxy and structure formation theories since there is substantial rate of creation of matter from dark energy, at the early epochs. The conversion of dark energy into matter at the sites of galaxy formation may even speed up the growth of matter density perturbation. Another novelty  is that the creation events may occur stochastically, so that the expansion rate of the universe, characterized by the Hubble parameter,  may become stochastic. This potentially can offer an explanation to the observed anomalous scatter in the Hubble diagram at high redshifts.

Undoubtedly, these premises   require a paradigm shift either in the galaxy  and  structure formation theories or in the evolution history in the early universe. Observations made in the unexplored territories of the universe often hold surprises and  our theories must be able to explain new data when it comes in.   There is lot of space in between  the farthest individually observed objects  and the cosmic microwave background.   The discovery of early bright galaxies with JWST is one such incident which demands modifications in  the cosmic dawn scenario.

\end{document}